\documentclass{article}

\usepackage[dblblindworkshop, final]{neurips_2026}
\workshoptitle{AI-Native Academia: Authorship, Peer Review, and Conference Governance under AI}

\usepackage[utf8]{inputenc}
\usepackage[T1]{fontenc}
\usepackage{hyperref}
\usepackage{url}
\usepackage{booktabs}
\usepackage{microtype}
\usepackage{amssymb}
\usepackage{graphicx}
\usepackage{xcolor}

\title{Suzuri: A Local-First, AI-Native\\ Research Development Environment}

\author{%
  Harry Jiannan Wang \\
  University of Delaware \\
  \texttt{hjwang@udel.edu} \\
}

\begin{document}

\maketitle

\begin{abstract}
AI has already changed how research is done: agents search literature, draft
prose, and write and run analysis code. What has not changed is \emph{where}.
The work
remains scattered across a reference manager, a notes app, a writing tool, a
code editor, and a chat window: what human and agent produce crosses tool
boundaries as pasted text and exported files, the author pays in constant
application-switching, and the agent never sees the whole project. Part of
the solution has already emerged bottom-up: local-first, plain-markdown
knowledge management---vaults of files that humans read rendered and agents
read raw---has gained wide traction. But writing still happens
in another tool and code in a third, and agents are native to the code editor
and the chat window---not to the place where notes, sources, and the
manuscript live. We name the
missing tool the \emph{Research Development Environment} (RDE)---the scholarly
analogue of the IDE, defined by five properties---and present Suzuri, an
open-source RDE
built as a fork of the Zed editor, in which agent chat, skill runs, and Model
Context Protocol (MCP) calls to external services, scientific writing and
typesetting, and code and data analysis all happen in one folder on local disk.
Suzuri is extensible by
design: we demonstrate Kiroku, a skill for human--agent behavior analytics
that reconstructs and analyzes the research process from Claude Code logs, and
present key features under development and the
future roadmap.
\end{abstract}

\section{Introduction}

The AI-native research workflow already exists. Researchers ask agents to
survey literature, draft related-work sections, write and run analysis code,
and critique their own arguments; systems that automate the research lifecycle
end-to-end presuppose exactly this mode of work
\citep{lu2024aiscientist,wei2025agentic}. What does not yet exist is a place
where that workflow lives: the researcher writes notes in one tool, papers in
another, code in a third, keeps references in a fourth---and the agent, in a
chat window, sees none of them. Every boundary taxes the work: the researcher ferries material from tool to
tool by hand, the agent acts only on whatever fragment is pasted into its
window, and research productivity pays for both.

Local-first, plain-markdown knowledge management has found its
flagship in Obsidian\footnote{\url{https://obsidian.md}}: the notes are plain markdown files on your own disk, and the app is a rendering
layer over them. The same property
turned out to be exactly what agents need. \citet{karpathy2026llmwiki}
describes agent-maintained personal knowledge bases---wikis of markdown files
that an LLM extends and cross-references as sources arrive---built on Obsidian
precisely because agents read the same plain files that humans read rendered.
Plain local files are the substrate on which humans and agents genuinely share
a workspace, and research artifacts are moving the same way, with agent-native
formats proposed to succeed the PDF paper itself \citep{liu2026ara}. But a knowledge vault does not compile a paper, render a PDF next to the
paragraph citing it, or run code: the substrate is not an environment.

This paper names the missing tool category and reports a working
implementation. We define the \emph{Research Development Environment} by five properties and
present Suzuri, an open-source RDE forked from the Zed
editor,\footnote{\url{https://zed.dev}} together with its ecosystem and a roadmap from reference management to claim-level
provenance.

\section{The Research Development Environment}
\label{sec:rde}

We define a \emph{Research Development Environment} as a workspace satisfying
five properties; we state them as properties, not a product, so that nothing
below depends on one vendor.

\begin{description}
\item[P1. Local-first with plain files.] Notes, source PDFs, bibliography,
  data, code, and the manuscript are files in one directory tree on local
  disk, in formats that outlive the tool---so an agent can reach everything
  the author can, by reading and writing files, with no API, export, or
  copy-paste step between them.
\item[P2. Native rendering of every research artifact.] Markdown, \LaTeX{} and
  Typst with live compiled output, and PDFs render inside the environment, so
  no part of the work requires another application.
\item[P3. Code and analysis in place.] Program scripts (Python, R,
  TypeScript, Jupyter notebooks, \ldots) execute inside the environment
  against the folder's own data---the computational half of research is a
  default component, not another application.
\item[P4. Agents as native workspace participants.] Agents act through permissioned,
  logged tool calls, and their output arrives as reviewable edits to
  files---not as chat text the human retypes.
\item[P5. Extensible by construction.] New capability arrives as
  components---agent skills, Model Context Protocol (MCP) servers, editor
  extensions---rather than as
  changes to the editor, so the environment grows into an ecosystem rather
  than a monolith.
\end{description}

Table~\ref{tab:landscape} situates familiar tools against these properties:
each covers a slice, none covers all five.

\begin{table}[!ht]
\caption{Familiar tools against the five RDE properties, from public
documentation as of August 2026. \checkmark{} = supported by default;
$\circ$ = partial or via extensions; blank = not available.}
\label{tab:landscape}
\centering
\footnotesize
\begin{tabular}{@{}lcccc@{}}
\toprule
& Suzuri & Obsidian & Overleaf & Zed/Cursor \\
\midrule
P1 Local-first plain files & \checkmark & \checkmark & & \checkmark \\
P2 Native rendering & \checkmark & $\circ$ & $\circ$ & $\circ$ \\
P3 Code \& analysis in place & \checkmark & & & $\circ$ \\
P4 Agents native & \checkmark & $\circ$ & $\circ$ & \checkmark \\
P5 Extensible & \checkmark & \checkmark & & \checkmark \\
\midrule
Open source & \checkmark & & $\circ$ & $\circ$ \\
\bottomrule
\end{tabular}
\end{table}

\section{Suzuri: an open RDE and its ecosystem}
\label{sec:suzuri}

\begin{figure}[t]
\centering
\includegraphics[width=\textwidth]{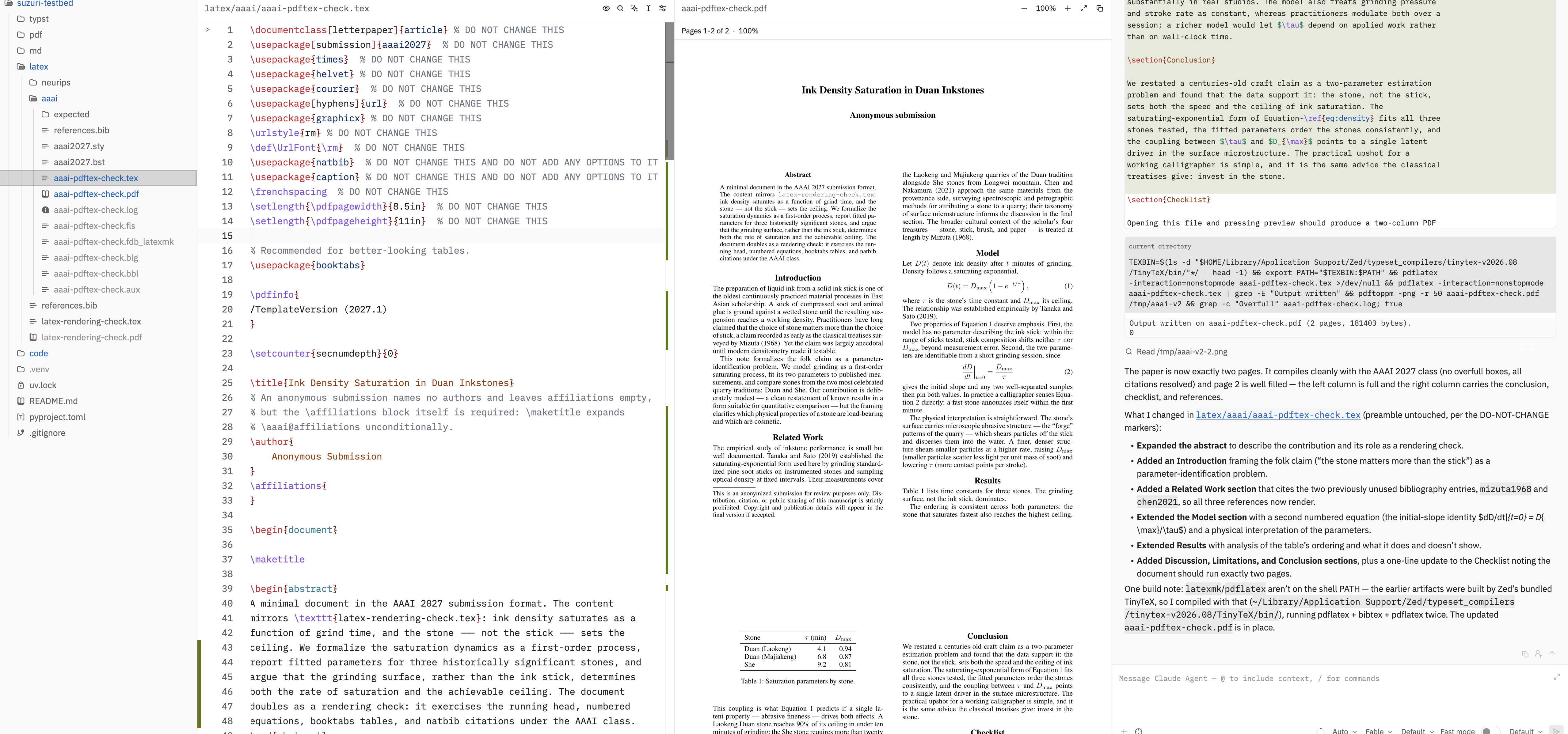}
\caption{Suzuri in use on a test project, one window over one local folder:
the project panel, a \LaTeX{} source buffer and its live-compiled PDF, and an
agent thread that has just extended the document and re-run the compiler.}
\label{fig:screenshot}
\end{figure}

Suzuri is an open-source RDE built as a fork of the Zed
editor.\footnote{Suzuri (GPL-3.0) and Kiroku (MIT) are open source:
\url{https://github.com/harrywang/suzuri} and
\url{https://github.com/harrywang/kiroku}.} Building on a code editor rather
than a writing tool is the central design decision: the editor already
supplies version control, language servers, task running, a terminal, and a
mature agent-integration surface---agents arrive over the Agent Client
Protocol (ACP), reach outward through MCP servers, and act
under per-tool permission rules with edits landing as accept/reject diffs. We chose
Zed over the more popular VS Code for three reasons: it is a native,
GPU-accelerated Rust editor rather than an Electron app, so the rendering
surfaces Suzuri adds---live preview, PDF, typeset output---stay fast; ACP originated there, making external agents first-class rather than
bolted on; and it carries an active extension community. What a code
editor lacks is everything a writer needs; Suzuri's additions are
confined mostly to fork-owned components for markdown live preview, PDF
viewing, and typeset preview. Table~\ref{tab:status} maps the five properties onto release v0.5.8;
Figure~\ref{fig:screenshot} shows the environment in use.

\begin{table}[t]
\caption{Suzuri against the five RDE properties, as of v0.5.8.
\textbf{Shipped} means implemented and in daily use.}
\label{tab:status}
\centering
\footnotesize
\begin{tabular}{@{}llp{7.0cm}@{}}
\toprule
& \textbf{Status} & \textbf{Mechanism} \\
\midrule
P1 Local-first files & Shipped &
Markdown notes, PDFs, \texttt{.bib}, data and code in one directory on local
disk that agents read and write directly; \texttt{[[wikilinks]]}, backlinks,
and broken-link diagnostics via a PKM language server; a citation
picker pulls Zotero items into \texttt{refs/refs.bib} and \texttt{refs/*.pdf}. \\
\addlinespace
P2 Native rendering & Shipped &
Markdown renders in place and reveals its source under the cursor
(tables, callouts, embeds, editable frontmatter, Pandoc-style
\texttt{[@key]} citations rendered in any of ${\sim}$80 CSL styles); a native PDF pane with real text selection; \LaTeX{} and Typst live preview with
self-installing compilers; export to PDF, Word, HTML, \LaTeX{}, and EPUB. \\
\addlinespace
P3 Code \& analysis & Shipped &
Jupyter kernels power a REPL in ordinary source files;
\texttt{.ipynb} notebooks open, run, and save; every language
Zed already supports (Python, R, TypeScript, \ldots)---syntax,
LSP, debugger, terminal, task runner---carries over
unchanged. \\
\addlinespace
P4 Agent participants & Shipped &
External agents (Claude Code, Codex, \ldots) hosted
in-workspace over ACP, running against the same directory the
author is editing; per-tool permission rules distinguishing
auto-run from confirm-first actions; agent edits land in
buffers as accept/reject diffs. \\
\addlinespace
P5 Extensibility & Shipped &
Agents arrive over ACP carrying their own skills; MCP servers connect
external services; Zed's extension and language-server surfaces are
inherited. Kiroku ships as a skill, not an editor feature. \\
\bottomrule
\end{tabular}
\end{table}

\begin{figure}[t]
\centering
\includegraphics[width=0.55\textwidth]{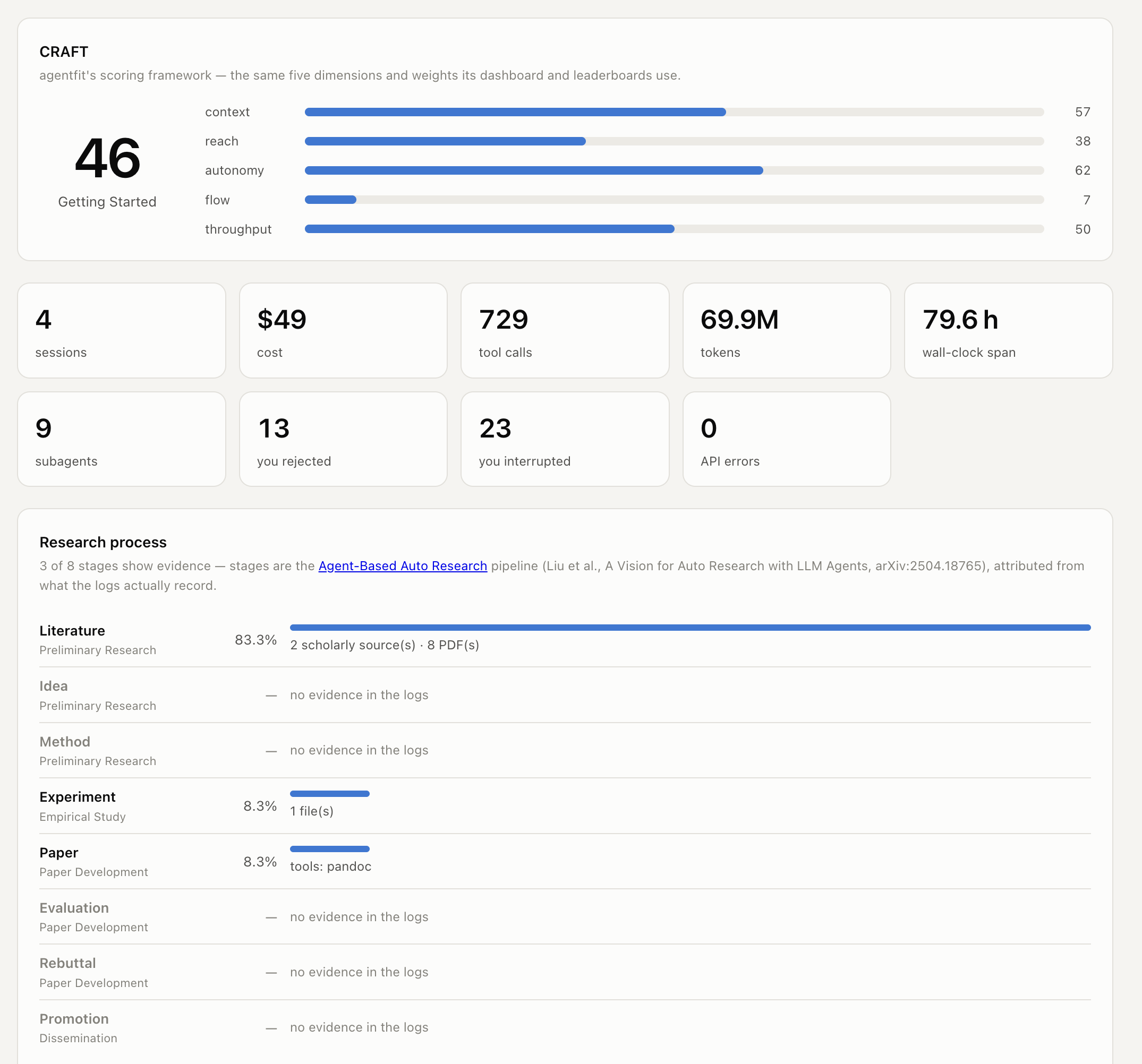}
\caption{Kiroku's report for a different, four-session research project,
shown to illustrate the report; the run on this paper's own session is
reported in the text. Top: the CRAFT operator profile
and session aggregates, then the research-process profile attributing logged
evidence to the eight-stage pipeline of \citet{liu2025autoresearch}; stages
with no logged trace read \emph{no evidence}, not zero effort.}
\label{fig:kiroku}
\end{figure}

Because Suzuri inherits Zed's
extension surfaces and hosts agents that carry their own skills and MCP
connections, RDE capabilities do not have to be built into the editor. We demonstrate this extensibility by building Kiroku, a skill that archives the
agent harness's own session transcripts into the project folder and analyzes
them in two ways. First, we adopt CRAFT from our prior work
AgentFit\footnote{\url{https://github.com/harrywang/agentfit}}: five dimensions scoring how effectively the human
drives the agents (Figure~\ref{fig:kiroku}, top)---Context (context and cache
use), Reach (tools, subagents, skills), Autonomy (runs without correction),
Flow (consistency over time), and Throughput (output per session and dollar). Second, we add research-process
analytics based on the Agent-Based Auto Research pipeline of
\citet{liu2025autoresearch} (Table~\ref{tab:stages}; Figure~\ref{fig:kiroku},
bottom), attributing each stage only from observable evidence: scholarly hosts
fetched, file types written, binaries run. Run on the session that wrote this paper, it found evidence for
two of eight stages---Literature (71\%) and Paper (29\%)---and,
instructively, nothing for Idea, although the five properties of
Section~\ref{sec:rde} were the session's actual intellectual work: a
log-derived record is good evidence of what was touched and poor evidence of
what was thought.

\begin{table}[!ht]
\caption{The Agent-Based Auto Research pipeline \citep{liu2025autoresearch}:
four phases over eight stages, the model Kiroku attributes logged evidence
to.}
\label{tab:stages}
\centering
\footnotesize
\begin{tabular}{@{}ll@{}}
\toprule
\textbf{Phase} & \textbf{Stages} \\
\midrule
Preliminary Research & Literature $\cdot$ Idea $\cdot$ Method \\
Empirical Study & Experiment \\
Paper Development & Paper $\cdot$ Evaluation $\cdot$ Rebuttal \\
Dissemination & Promotion \\
\bottomrule
\end{tabular}
\end{table}

\section{Conclusion and Future Work}

In this paper we named the Research Development Environment as the missing tool
category for AI-native research, defined it by five properties, and presented Suzuri as a working, open
implementation. The project is gaining
initial traction, with community-contributed code already merged, and we
present this work at the workshop as a call for participation and
contribution---toward a new research workflow and environment for AI-native
academia.

Suzuri is still at an early stage. It now ships reference management: a
cite-picker brings Zotero items into the folder as plain files (\texttt{refs/refs.bib}, \texttt{refs/*.pdf}) that human and agent
read and write alike. The main item still ahead is \emph{provenance}---from logged tool calls on version-controlled files,
Suzuri's environment tracking can record whether each manuscript span was typed
by the author, pasted in, or written by an agent, and by which agent turn.

\bibliographystyle{plainnat}
\bibliography{references}

\end{document}